\documentclass[english]{article}

\usepackage[T1]{fontenc}
\usepackage[latin9]{inputenc}
\usepackage{verbatim}
\usepackage{amsmath}
\usepackage{amssymb}
\usepackage{graphicx}

\makeatletter
\newcommand{\code}[1]{\texttt{#1}}

\makeatother

\begin{document}

\title{Category Theory for Modeling OOP\footnote{This extended abstract, in abridged form, has been accepted for presentation at The Nordic Workshop on Programming Theory (NWPT'17), Turku, Finland, November 1-3, 2017.}}

\author{Moez A. AbdelGawad\\
\medskip{}Informatics Research Institute, SRTA-City, Alexandria, Egypt\\
\code{moez@cs.rice.edu}}
%\institute{Informatics Research Institute, SRTA-City, Alexandria, Egypt\\
%\email{moez@cs.rice.edu}}
%\authorrunning{Moez AbdelGawad}
%\titlerunning{Category Theory for Modeling OOP}

\maketitle
Category theory has been used extensively and successfully in modeling
functional programming languages (see, \emph{e.g.},~\cite{Scott82,CatRecDomEqs82,GunterHandbook90,Pierce91,spivak2014category,DomTheoryIntro}).
However, it has been used to a lesser extent in understanding and
modeling object-oriented programming (OOP) languages, mainly focusing on OO languages extant during
the early days of OOP research~\cite{CanningFbounded89,CookInheritance90,Jacobs95,Poll2000276}.

Recently, we presented a detailed outline for using operads, from
category theory, to model the iterative construction of the infinite
subtyping relation in Java and other generic nominally-typed OO programming languages
such as C\#, C++ and Scala. Besides using operads to model the construction
of the subtyping relation,
we believe that there are plenty of other new uses of category-theoretic
tools that can help in having better models and a better understanding
of mainstream OOP languages.

In this extended abstract we present outlines for four potential applications
of category theory in OOP research. Namely, we first present (1) a
summary of our use of operads to construct the Java subtyping relation,
then we present (2) the possible use of representable functors (and Yoneda's
Lemma) in modeling and understanding generic types of generic nominally-typed
OOP, followed by (3) the possible use of the equivalence of category
presentations to relational database schema and of cartesian-closed
categories as models of functional programming to model a structural view
of OOP, and, finally, we present (4) the possible use of adjoint functors to model
a particularly complex feature of Java generics, namely Java erasure.

\textbf{\emph{Operads and Generic OO Subtyping.}} \global\long\def\JSO{\mathcal{JSO}}
Earlier this year, in~\cite{AbdelGawad2017a,AbdelGawad2017}, we outlined how an
operad, called $\JSO$ (for Java Subtyping Operad), can be defined
to model the iterative construction of the generic subtyping relation
in Java and other similar generic nominally-typed OO languages such
as C\# and Scala. Our model makes use of two facts:
the fact that the generic subtyping relation in Java exhibits
intricate self-similarity,
due to the existence of wildcard types (and, accordingly,
the existence of three subtyping rules for generic types), and that
operads are category-theoretic tools suited for and frequently used
to model self-similar phenomena. (More on operads can be found in~\cite{spivak2014category,leinster2004higher}).

%\begin{comment}
\begin{figure}
\noindent \begin{centering}
\includegraphics[angle=270,scale=0.75]{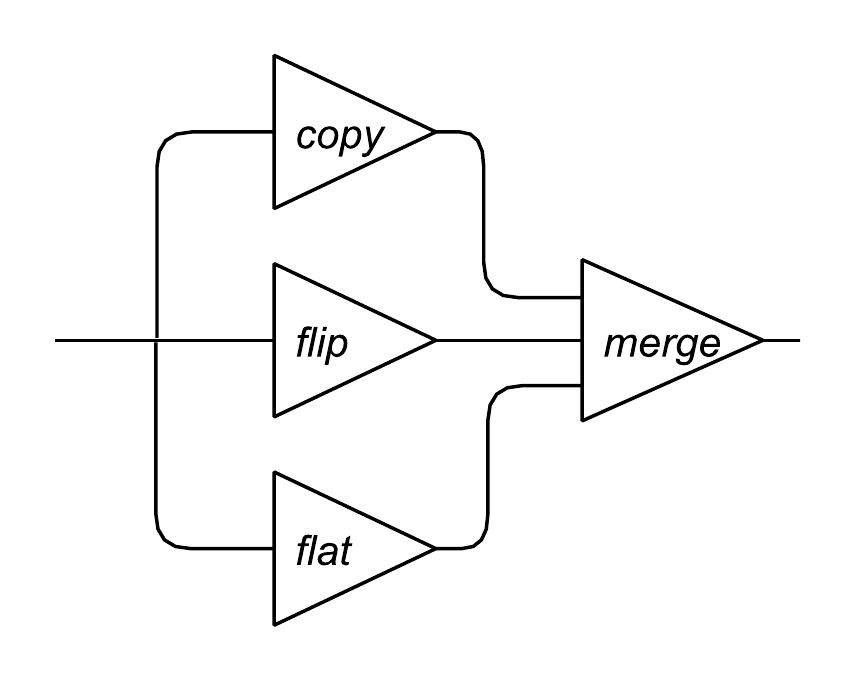}
\par\end{centering}

\protect\caption{\label{fig:JSM}The $JSM$ morphism of $\protect\JSO$}
\end{figure}
%\end{comment}
In $\JSO$ we present four operad morphisms, $copy$, $flip$, $flat$,
and $merge$ that model the three subtyping rules for generic types
in Java and a (hidden) merging transformation. The operadic composition of
these four morphisms (see
%\begin{comment}
Figure~\ref{fig:JSM})
%\end{comment}
%~\cite[Figure~3.1]{AbdelGawad2017a,AbdelGawad2017})
defines how the subtyping relation in Java can be iteratively constructed.
More details on $\JSO$, and examples that illustrate its inner workings
and its application, are found in~\cite{AbdelGawad2017a,AbdelGawad2017}.

Recently, while working on formalizing $\JSO$ in full, it has been
suggested to us~\cite{Spivak2017} to make the presentation of $\JSO$
clearer by specifying some of the ``obvious'' relations that exist
between morphisms of $\JSO$. As such, the most obvious of the relations
between $\JSO$ morphisms seem to us to be $copy=id$, $flip^2=flip\circ flip=copy$,
and $flatten^2=flatten\circ flatten=flatten$, which express that $copy$ is
an identity transformation, that $flip$ is its own inverse, and that
$flatten$ is an idempotent transformation. We intend to present more
of these relations (and more accurate versions of them) when, in future
work, we present the full formal definition of $\JSO$.

Using operads to model the subtyping relation in Java has revealed
to us some facts that we believe improve our understanding of the
type systems of generic nominally-typed OO languages. These include,
for example, revealing three reasons why the self-similarity of the
Java subtyping relation is intricate and not immediately obvious (see~\cite[Section~6]{AbdelGawad2017a,AbdelGawad2017}).
It has also made clearer to us some suggestions on how to build a
domain-theoretic or categery-theoretic model of generic nominally-typed
OO languages, as we explain below.

\textbf{\emph{Yoneda's Lemma, Representable Functors, and OO Generic
Types.}} Last year we developed and presented few ideas on how to
construct a domain-theoretic model of generic nominally-typed OOP~\cite{AbdelGawad2016a}
along similar lines to our earlier construction of a model of non-generic
nominally-typed OOP~\cite{NOOP,NOOPbook,NOOPsumm}, particularly
suggesting using the notions of \emph{nominal type intervals} and
\emph{full generification} and illustrating these notions with a number
of code examples~\cite{AbdelGawad2016c}. Recently we found out that,
again, category theory can be employed to better improve, explain and present our model.

David Spivak has presented, in his famed book~\cite{spivak2014category}
and also earlier in~\cite{Spivak2012}, how to view relational database
schemas as category presentations and vice versa. In his work, Spivak
also explains the meaning of representable functors, and of the related Yoneda's
lemma, in the context of relational databases, where representable
functors are viewed as particular (``most generic'') database instances
(of a database schema) and Yoneda's lemma provides a correspondence
between each row of a relational database table and an instantiation
of the most generic instance of the the database schema corresponding
to that table. 

Given that representable functors can be viewed as modeling the \emph{most}
\emph{generic} instances of objects in a category and that Yoneda's
lemma states that specifying an instance of some type is the same
as `filling in all placeholders' in the most generic instance of
an object~\cite{spivak2014category}, it immediately came to our
minds that both notions can be helpful in understanding and modeling
generic\emph{ }nominally-typed OOP. %

In particular, we saw that if generic classes of a Java program are objects of a category where non-id arrows of the category connect a generic class to classes of the upperbounds (\emph{i.e.}, the erasure) of its type parameters (classes that take no type parameters will have no non-id outgoing arrows), then in that context Yoneda's lemma is simply a formal expression of the fact that defining and specifying a generic type (in Java and other similar languages) is done just by specifying a generic class and its actual type arguments (\emph{e.g.}, for generic class \code{List} and actual type argument \code{String}, we get type \code{List<String>}).

Also, given a database schema/category $C$, then just as Spivak uses set-valued functors $F: C\rightarrow Set$ to model instances of $C$, we found that set-valued functors can similarly model generic types as instances (also called `instantiations') of generic classes, where a generic class gets mapped by such a functor to the corresponding set of its instantiations. Further, given that not all set-valued functors are representable functors, but that representable functors are rather the most generic
instances/functors, we found that representable functors over the
category of generic classes of a program can correspond to instantiations
of the \emph{fully-generified} versions of the classes---which we explain in~\cite{AbdelGawad2016c}---using artificial type variables (equivalent to
Skolem variables in logic, and to labelled nulls in database systems), making clearer a connection between our own work and category theory.

\textbf{\emph{SOOP=RDB+FP=Cat+CCC.}} Compared to a nominally-typed
view of OOP~\cite{NOOP,NOOPbook,NOOPsumm,InhSubtyNWPT13,AbdelGawad2015},
a structurally-typed view of OOP is simpler, but also less accurate.
A structural view of OOP is still held by some OO PL researchers~\cite{OOPOverview13,DomThSummCOOP14,AbdelGawad2016,AbdelGawad2015a}.
A structural view of OOP views objects as being mere records, of data
values (as fields) and of functions (as methods), \emph{i.e.}, as records that
do \emph{not} include nominal class/type information that is associated with \emph{behavioral contracts} that the objects
maintain~\cite{AbdelGawad2015}. Due to its semantic significance, the inclusion of nominal information is essential in the nominally-typed view of OOP (where it is usually included as a third class signature component of objects). In the simpler structural
view of OOP an object can however be viewed as being composed of just two components: an
inactive ``state'' component (fields of the object, or its ``configuration'')
and an active ``processing'' component (methods of the object, or
its ``actions'').

As hinted to earlier, Spivak has shown the equivalence of relational
database schemas and category presentations. Also, it is well-known
that functional programming, and typed $\lambda$-calculus (the ``mother''
of all functional programming languages) in particular, can be modeled
by \emph{cartesian closed categories }(\emph{CCCs})\emph{.}

Given the strong similarity between fields of objects and data fields
in a relational database table and the strong similarity between methods of
objects and (mutually-recursive) functional programming, we conjecture
that combining both models of relational databases and functional
programming in some precise way can be used to model structural OOP.
This seems to be a plausible category-theoretic view of structurally-typed
OOP~\cite{Spivak2017}. In particular, we believe an accurate category-theoretic model of (structural) OOP can be developed that uses (instances
of) a category presentation to model fields of objects and (instances
of) cartesian closed categories to model methods of objects. In other
words, we believe structural OOP can be modeled by combining the category
theory models of relational databases (namely, category presentations),
and of functional programming (namely, cartesian closed categories).

In summary, we thus believe
\[
SOOP=Cat+CCC
\]
where $SOOP$ stands for a categorical model of structural OOP, $Cat$
is the category of (small) categories (or, more accurately, their
presentations), $CCC$ is the category of cartesian
closed categories, and $+$ is a
combination/composition operator whose details are so far left unspecified.

To model nominally-typed OOP using similar category-theoretic tools, it seems that including nominal information (as the third class signature component of objects) can be done via an algebra or an algebra signature, but we have not investigated this idea further. We leave it to future work.

\textbf{\emph{Adjoint Functors and Java Erasure.}} %
Adjoint functors (a pair of which define an adjunction) can
be best understood intuitively as modeling approximations, or as providing so-called
``formulaic optimization''. For example, the floor and ceiling functions
(usually denoted by $\left\lfloor \cdot\right\rfloor $, $\left\lceil \cdot\right\rceil $),
which map real numbers in $\mathbb{R}$ to their nearest integers in $\mathbb{Z}$, can be presented
as examples of two adjunctions when each of the two functions is separately paired with the
inclusion function $i:\mathbb{Z\rightarrow\mathbb{R}}$ that maps
integers into the real numbers. The floor and ceiling of a real number
present a best-lower and best-upper approximation of that real number.
A similar sense of approximation exists in Java erasure, where a (less
precise) non-generic type can be viewed as approximating a (more precise)
generic type.

The current support for generics in Java depends heavily on the notion
of Java erasure, which is a feature of Java that eases `migration-compatibility'
from non-generic Java code to generic Java code. Java erasure ``erases''
a generic type to a related non-generic type, which also allows generic
code to be run on non-generic JVMs. (For example, instantiations of
the generic Java class \code{LinkedList}, such as \code{LinkedList<String>}
and \code{LinkedList<Integer>}, get erased to the non-generic legacy
Java type \code{LinkedList}.)

Given the one-to-one correspondence between non-generic subtyping
and subclassing (particularly type inheritance) in nominally-typed OO languages
such as Java, C\# and Scala~\cite{InhSubtyNWPT13,AbdelGawad2015a},
erasure can be viewed as a mapping from generic types to Java classes,
where the subclassing relation between generic classes/type constructors
is \emph{the same} as the non-generic subtyping relation. Erasure,
as such, defines an (adjoint) functor, say $F$, from the generic
subtyping relation to the subclassing relation. The other element
in the pair of functors (so as to define an adjunction) should be a functor,
say $G$, that takes a generic class and produces a ``canonical''
instantiation of that class. We take $G$ to be the functor instantiating
the generic class with the wildcard type `\code{?}' (which by default
is bounded by the \emph{upperbound} of its corresponding type parameter.
With little difficulty, F-bounded type parameters, where a type parameter
is used in its own bound, can be handled by $G$).

To demonstrate that $F$ and $G$ define an adjunction (called a `Galois
connection' in the context of partial orders, since the generic subtyping
relation and the subclassing relation are indeed partial orders),
we should note that it is easy to see that the condition for an adjunction
is satisfied by $F$ and $G$. Namely, in Java we have 
\[
F(a)\le b\Longleftrightarrow a<:G(b),
\]
 for all generic types $a$, generic classes $b$, where $\le$ denotes
the subclassing relation and $<:$ denotes the generic subtyping relation.

In words, this condition says that the erasure $F(a)$ of a generic type $a$
 is a subclass of class $b$ if and only if
$a$ is a subtype of the `\code{?}' wildcard instantiation $G(b)$ of class $b$, which is a true statement in Java and all generic nominally-typed
OO languages. (To illustrate, consider, for example, the statement
\[
\mathtt{LinkedList\le List\Longleftrightarrow LinkedList\negthickspace<\negthickspace T\negthickspace>\;<:\;List\negthickspace<?\;extends\;Object\negthickspace>}
\]
 where $a$ is the generic type \code{LinkedList<T>} for any type
parameter $\mathtt{T}$---\emph{e.g.}, \code{String} or \code{Integer}
or `\code{?~extends Number}'---$b$ is class \code{List}, and \code{Object}
is the upperbound of the type parameter of class\footnote{In this work we treat Java interfaces as abstract classes.}
\code{List}. This statement asserts that \code{LinkedList} is a
subclass of \code{List} if and only if any instantiation of \code{LinkedList}
is a subtype of \code{List<?~extends Object>}---the same
type as \code{List<?>}---which is a true statement in Java.)

Given that any adjunction induces a monad, and vice versa, to strengthen the
connection with category-theoretic notions even further we can
present the monad induced by the $(F,G)$ adjunction, but for space
considerations we refrain from doing so here. In summary, though, we should note that, in Java, for all generic types $a$ and generic classes $b$, we have
\[
a <: G(F(a))\mathrm{~and~}F(G(b)) \le b
\]
(in fact, more strongly, in Java we have $b=F(G(b))$).\\

%The following paragraph was NOT included in the final version submitted to NWPT'17, and was added here afterwards.
Of course, it is natural to expect that there could be strong connections and interactions between the four uses of category theory we presented above. For example, note the use of Java erasure when discussing representable functors and the application of Yoneda's lemma to the category of generic classes.  Also, note that Yoneda's lemma will be applicable ``at two different levels'' (\emph{i.e.}, the level of values and the level of types) if nominally-typed OOP is modeled as we suggested above (it will applicable to the instantiation of generic classes to define generic types, and to the instantiation of generic types to define objects, \emph{a.k.a.}, instances of these types).  We keep making further connections to future work, however.

To conclude this abstract, based on the connections we made above between some notions of category theory and some features of OOP languages, particularly generic nominally-typed OOP languages, we believe that the extensive use of category theory in modeling functional programming two decades ago should now be matched by more use of category theory---including the use of its newly discovered tools---in improving our understanding of modern mainstream OOP languages.

\bibliographystyle{plain}
%\bibliography{D:/Work/NOOP}

\begin{thebibliography}{10}
	
	\bibitem{NOOP}
	Moez~A. AbdelGawad.
	\newblock {\em {NOOP}: A Mathematical Model of Object-Oriented Programming}.
	\newblock PhD thesis, Rice University, 2012.
	
	\bibitem{NOOPbook}
	Moez~A. AbdelGawad.
	\newblock {\em {NOOP}: A Nominal Mathematical Model Of Object-Oriented
		Programming}.
	\newblock Scholar's Press, 2013.
	
	\bibitem{OOPOverview13}
	Moez~A. AbdelGawad.
	\newblock An overview of nominal-typing versus structural-typing in
	object-oriented programming (with code examples).
	\newblock Technical report, {arXiv.org:1309.2348 [cs.PL]}, 2013.
	
	\bibitem{NOOPsumm}
	Moez~A. AbdelGawad.
	\newblock A domain-theoretic model of nominally-typed object-oriented
	programming.
	\newblock {\em Electronic Notes in Theoretical Computer Science (ENTCS)},
	301:3--19, 2014.
	
	\bibitem{DomThSummCOOP14}
	Moez~A. AbdelGawad.
	\newblock Domain theory for modeling {OOP}: A summary.
	\newblock Technical report, {arXiv.org:1406.7497 [cs.PL]}, 2014.
	
	\bibitem{AbdelGawad2016}
	Moez~A. AbdelGawad.
	\newblock A comparison of {NOOP} to structural domain-theoretic models of
	object-oriented programming.
	\newblock {\em Preprint available at http://arXiv.org/abs/1603.08648}, 2016.
	
	\bibitem{AbdelGawad2016c}
	Moez~A. AbdelGawad.
	\newblock Towards an accurate mathematical model of generic nominally-typed
	{OOP} (extended abstract).
	\newblock {\em {arXiv:1610.05114 [cs.PL]}}, 2016.
	
	\bibitem{AbdelGawad2016a}
	Moez~A. AbdelGawad.
	\newblock Towards understanding generics.
	\newblock Technical report, {arXiv:1605.01480 [cs.PL]}, 2016.
	
	\bibitem{AbdelGawad2015}
	Moez~A. AbdelGawad.
	\newblock Why nominal-typing matters in {OOP}.
	\newblock {\em Preprint available at http://arxiv.org/abs/1606.03809}, 2016.
	
	\bibitem{AbdelGawad2017a}
	Moez~A. AbdelGawad.
	\newblock Towards a {J}ava subtyping operad.
	\newblock {\em Proceedings of FTfJP'17, Barcelona, Spain, June 18-23, 2017},
	2017.
	
	\bibitem{AbdelGawad2017}
	Moez~A. AbdelGawad.
	\newblock Towards a {J}ava subtyping operad (extended version).
	\newblock {\em Preprint available at http://arxiv.org/abs/1706.00274}, 2017.
	
	\bibitem{AbdelGawad2015a}
	Moez~A. AbdelGawad and Robert Cartwright.
	\newblock In nominally-typed {OOP}, objects are not mere records and
	inheritance \emph{Is} subtyping.
	\newblock {\em Submitted for journal publication}, 2016.
	
	\bibitem{CanningFbounded89}
	Peter~S. Canning, William~R. Cook, Walter~L. Hill, J.~Mitchell, and W.~Olthoff.
	\newblock F-bounded polymorphism for object-oriented programming.
	\newblock In {\em Proc. of Conf. on Functional Programming Languages and
		Computer Architecture}, 1989.
	
	\bibitem{InhSubtyNWPT13}
	Robert Cartwright and Moez~A. AbdelGawad.
	\newblock Inheritance \emph{Is} subtyping (extended abstract).
	\newblock In {\em The 25\textsuperscript{th} Nordic Workshop on Programming
		Theory (NWPT)}, Tallinn, Estonia, 2013.
	
	\bibitem{DomTheoryIntro}
	Robert Cartwright, Rebecca Parsons, and Moez~A. AbdelGawad.
	\newblock {\em Domain Theory: An Introduction}.
	\newblock Preprint available at http://arxiv.org/abs/1605.05858, 2016.
	
	\bibitem{CookInheritance90}
	William~R. Cook, Walter~L. Hill, and Peter~S. Canning.
	\newblock Inheritance is not subtyping.
	\newblock In {\em POPL'90 Proceedings}, 1990.
	
	\bibitem{GunterHandbook90}
	C.~A. Gunter and Dana~S. Scott.
	\newblock {\em Handbook of Theoretical Computer Science}, volume~B, chapter 12
	(Semantic Domains).
	\newblock 1990.
	
	\bibitem{Jacobs95}
	Bart Jacobs.
	\newblock Objects and classes, coalgebraically.
	\newblock In {\em Object-Orientation with Parallelism and Persistence}, pages
	83--103. Kluwer Acad. Publ, 1996.
	
	\bibitem{leinster2004higher}
	T.~Leinster.
	\newblock {\em Higher Operads, Higher Categories}.
	\newblock London Mathematical Society Lecture Note Series. Cambridge University
	Press, 2004.
	
	\bibitem{Pierce91}
	Benjamin~C. Pierce.
	\newblock {\em Basic Category Theory for Computer Scientists}.
	\newblock MIT Press, 1991.
	
	\bibitem{Poll2000276}
	Erik Poll.
	\newblock A coalgebraic semantics of subtyping.
	\newblock {\em Electronic Notes in Theoretical Computer Science}, 33:276 --
	293, 2000.
	\newblock CMCS'2000, Coalgebraic Methods in Computer Science.
	
	\bibitem{Scott82}
	Dana~S. Scott.
	\newblock Domains for denotational semantics.
	\newblock Technical report, Computer Science Department, Carnegie Mellon
	University, 1983.
	
	\bibitem{CatRecDomEqs82}
	M.~B. Smyth and Gordon~D. Plotkin.
	\newblock The category-theoretic solution of recursive domain equations.
	\newblock {\em {SIAM} J. of Computing}, 11:761--783, 1982.
	
	\bibitem{Spivak2012}
	David Spivak.
	\newblock Functorial data migration.
	\newblock {\em Information and Computation}, 217:31--51, August 2012.
	
	\bibitem{spivak2014category}
	David Spivak.
	\newblock {\em Category theory for the sciences}.
	\newblock MIT Press, 2014.
	
	\bibitem{Spivak2017}
	David Spivak.
	\newblock Private communication.
	\newblock 2017.
	
\end{thebibliography}

\end{document}